# Using Steganography and Watermarking
# For Medical Image Integrity

By Givon Zirkind

HZ85@mynova.nsu.edu

ISEC 0620 Fall 2024

# ABSTRACT

Medical imaging has kept up with the digital age. Medical images such as x-rays are no longer keep on film or; even made with film. Rather, they are digital[1] [2]. In addition, they are transmitted for reasons of consultation and telehealth as well as archived. Transmission and retrieval of these images presents an integrity issue, with a high level of integrity being needed. Very small artifacts in a digital medical image can have significant importance, making or changing a diagnosis. (Magdy, et. al., 2022) (Kahlia, et. al. 2023) It is imperative that the integrity of a medical image, especially in a Region of Interest (ROI) be identifiable and preserved. (Badshah, et. al. 2016) Watermarking and steganography are used for the purposes of authenticating images, especially for copyright purposes. (Hamza et. al., 2021) These techniques can be applied to medical images. However, these techniques can interfere with the picture's integrity. While such distortion may be acceptable in other domains, in the medical domain this distortion is not acceptable. High accuracy is imperative for diagnosis. (Kahlia, et. al. 2023) This paper discusses the techniques used, their advantages and shortcomings as well as methods of overcoming obstacles to integrity.

---

[1] https://www.recordrs.com/blog/medical-digital-imaging-and-its-impact-on-medical-record-retrieval/
[2] http s://chartrequest.com/what-is-digital-medical-imaging/

# Table of Contents

# Table of Figures

# INTRODUCTION

Along with other advances in the digital age, medical imaging has gone from silver nitrate photographic images to digitized images. This impacts methods of retaining the integrity of the image, security and archival methods. As images are transmitted, there is a chance that there will be data loss. Small data loss, imperceptible to the human eye can cause misdiagnosis, especially when diagnosing with AI. (Lee et. al. 2023) AI has been proven capable of diagnosing what doctors can not and making far reaching predictions, which allows for current interventions to preclude disease outcomes. E.g. Image analysis of feet from diabetics can prevent amputations.[3] [4] [5] (Tehsin, et. al., 2023) To do this, minutiae of image detail must be present. If there are pixels missing or their values are altered, even ever so slightly, the AI calculations can be off, and misdiagnosis will occur. (Lee et. al. 2023)

Such losses can occur in image transmission. (Badshah, et. al. 2016) With the advent and promotion of telehealth, this is a much more serious issue. As (Tehsin, et. al., 2023) state, AI imaging to diagnose foot ulcers in diabetics can be used to reach those who do not live in proximity or have access to experts. Even within cities, Luminary Labs (See footnotes 1, 2 and 3.) has produced software with which diabetics can take pictures of their feet with their phones, then, receive an analysis, warnings and preventions, without a doctor. The outpatient's physician is then notified, and consultations can be arranged.

A new issue with telehealth and digital health imaging is medical identity theft. (Lee, et. al. 2023) A matter that can be addressed by steganography as well.

Steganography and watermarking techniques can be used to ensure the integrity of digital images. Steganography is the art or practice of concealing a message, image, or file within

---

[3] https://www.linkedin.com/pulse/finalists-presentation-alexa-diabetes-challenge-demo-day-zirkind
[4] https://aws.amazon.com/blogs/apn/announcing-the-alexa-diabetes-challenge-powered-by-luminary-labs/
[5] https://www.luminary-labs.com/open-innovation/alexa-diabetes-challenge/

another message, image, or file. (Dictionary, O. E. 1989) "The goal behind steganography techniques is to share data between the source and destination in an unnoticeable format." (Magdy, et. al., 2022)

Watermarking is a mark in the image that can not be removed from the image and does not detract from or degrade the image. This technique is commonly known and used as an anti-counterfeiting measure for currency.[6] The goal of watermarking is to ensure the authenticity of the artifact.

Both techniques can be applied to medical images to preserve integrity if safeguards are implemented to not prevent degradation of the integrity of the image. In combination with other techniques, confidentiality may also be provided for secure transmission of the images over the internet.

In this instance, attack refers to any factor that could degrade the image quality and alter the pixels. (Magdy, et. al., 2022) The effectiveness of the technique can be measured in a variety of ways. Testing the peak signal to noise ratio (PSNR) and comparing diagnostic output (from AI) from images with and without a steganographic application. (Magdy, et. al., 2022) The difference between the original picture and the picture after a steganographic transformation is considered noise. The goal being to keep the PSNR below a threshold for degrading the image as well as detecting a hidden message. (Aziz, et. al. 2020)

This paper will review steganographic and watermarking techniques in general. Then, discuss specific steganographic or watermarking techniques used for digital images. The techniques will be assessed for viability of use with medical images which includes: distortion, maintaining diagnosis and confidentiality.

Future research is proposed for methods not seen in the literature reviewed.

---

[6] https://www.americanbanker.com/news/watermarks-an-appreciation-for-a-timeless-feature-of-currency

# 1. Steganography

Steganography, also referred to as Reversible Data Hiding (RDH), is the science of communicating information while hiding its existence. This technique can be used for preserving the integrity of images, especially medical images. Steganographic techniques involve the embedding of one image inside another image. "This differs from cryptography, the art of secret writing, which is intended to make a message unreadable by a third party but does not hide the existence of the secret communication." (Kessler, et. al. 2004) Nevertheless, steganography can be considered a kind of cryptography. Steganography hides the message in a "medium". The message or "payload" is embedded in a medium, in this instance, the medical image. The size of the payload depends on the "capacity" of the medium. (This will be subsequently demonstrated.)

When using an image as the medium, the size of the image defines the medium's capacity. The image must have a capacity sufficient to hold the payload. This can create the need to use large images as mediums.

A common technique is to use a mask. The mask being a specific bit pattern that is XORed with the medium to hide the message in the medium. The mask is likened to an encryption key.

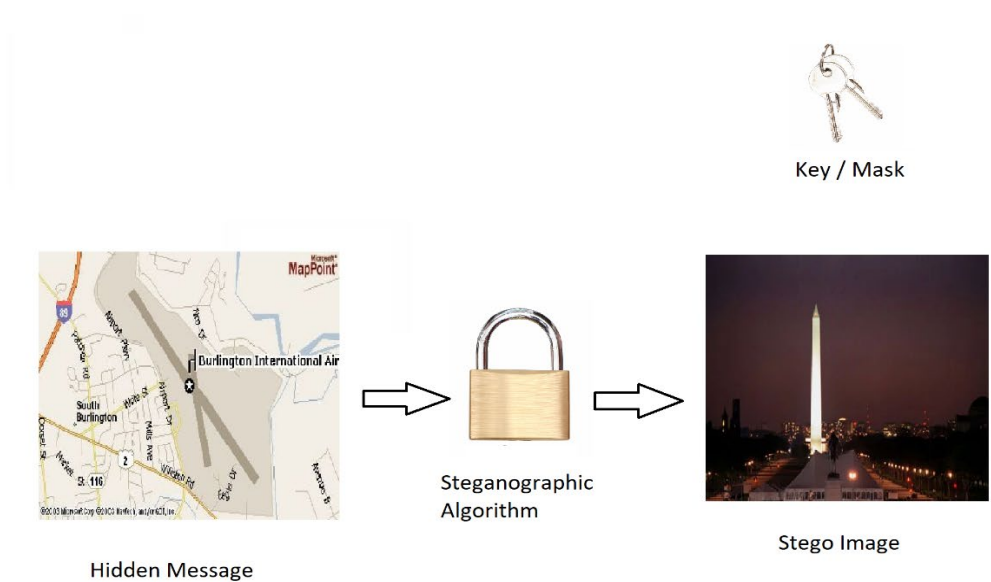

Figure 1. Basic Steganographic Process. Photos courtesy Kessler, GC (2004)

Lease Significant Bit (LSB) is one of the most commonly known and basic techniques for steganography. The message is embedded in the least significant bit of the pixels, usually following some order for selection of pixel bytes throughout the image. Hamza, et. al. 2021 states that between 1-4 bits can be used for LSB. For example, starting with the first byte and going on sequentially to the end of the file or a stop signal is reached. While hidden message may be laid out sequentially, the bits should be randomized to prevent easy detection. (Kessler, 2004) Encryption with a cryptographic method prior to embedding can provide this randomization of the bits.

This example demonstrates the need for a large carrier image, because each byte of a message will require 8 bytes in the carrier. For added secrecy, the order of the embedded bits can be scrambled and kept in a key. The key should be distributed separately from the steganographic image.

The author posits that using a key that is produced by a random number generator would increase the difficulty of extracting the hidden image. (In other work, the author has already demonstrated that using a key from a non-repeating, random sequence can produce a unique watermark.)

LSB works because human vision is insensitive to small changes in the image. (Wu, et. al. ,2003)

Chromatic steganography is an extension or premutation of the LSB be technique. Digital images can be encoded in several ways. One model is the Red Green Blue (RGB) model.[7] In this method, colors are expressed as triad of 3 hex numbers (00-FF) for RGB (#FFFFFF). The range of 00-FF indicating the intensity. For black/white, #000000 is black, #FFFFFF is white and #808080 is gray. Using this knowledge, similar to the LSB method, the hidden message can be in any one or combination of the color channels. While this does allow for a certain amount diffusion, the author has found in experimentation that relying on a single channel will cause color distortion. This vulnerability makes this method insecure without additional techniques to prevent distortion.

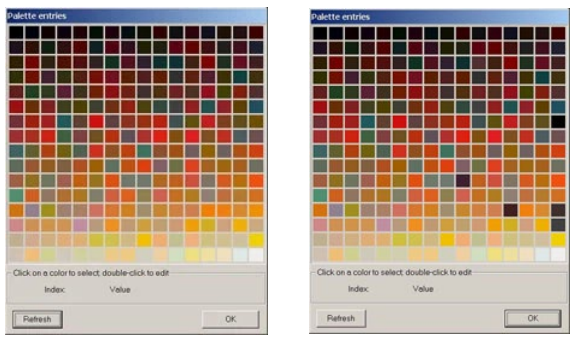

*Figure 2.* Comparison of color channels before/after steganographic embedding. One can see the emergence of black regions where embedding has taken place. Photos courtesy Kessler, GC (2004)

4LSB is least significant bit encoding that uses 4 significant bits. This gives 50% capacity but is prone to degradation. (Boryczka, et. al. 2023) This is the most prevalent technique used for steganography. A smaller number of bits can be used with a decrease in capacity.

---

[7] W3Schools, Colors Tutorial, https://www.w3schools.com/colors/default.asp

## 1.1. Other Considerations

In general, distortion of the image is a concern when using steganography, because too much distortion would raise suspicion of a hidden message. (Boryczka, et. al. 2023)

In an effort to increase the capacity of the medium or, conversely, to decrease the size of the payload, compression can be used. A technique we will soon explore.

Encryption of the payload could be used for confidentiality. So that even if the message is discovered, it is unreadable and undecipherable in its encrypted state. (Boryczka, et. al. 2023) Boryczka, et. al. 2023 suggest that the key can hidden with steganography to recover the encrypted message. If Boryczka, et. al. 2023 meant in a separate carrier, the author would agree. If within the same message, the author disputes that as a bad practice and obscurity security.

Just as there are ROIs (Regions Of Interest), there are also RONIs (Regions Of Non Interest). This is significant as to whether or not the distortion is tolerable or can be ignored. Should an image be altered it if it can be known that the altered region is a RONI, that would retain the usefulness of an image. A special consideration with medical images is that the edges may not be significant and even if distorted, the integrity of the image may be preserved. (Kahla, et. al., 2023)

There are 2 general kinds of digital steganography:

### Spatial Domain
Steganography in the spatial domain refers to the embedding of pixels and altering their bit values. The pixels of an image are considered as a 2D grid and modified directly. (Boryczka, et. al. 2023)

### Frequency Domain
Steganography in the frequency domain refers to embedding hidden information into the frequency of a cover image. E.g. The DCT (Discrete Cosine Transform) of a JPEG image.

Note, that both steganography and cryptography secure data transmission in public channels but, steganography tries to hide the fact that there is any communication at all.

With steganography, the concept of security (confidentiality) is very often used as one of the main criteria for its quality (of the image and imperceptibility of the message within the image. (Magdy, et. al. 2022)

## Peak Signal to Noise Ratio (PSNR)

PSNR (Peak Signal-to-Noise Ratio) measures the quality of the image subjected to steganographic or watermarking techniques. PSNR is calculated by comparing the original cover image to the steganographic image. Essentially, the difference of the pixel values from the original and steganographic image are summed and divided by the inverse of the total number of pixels. Thus, giving an average difference with which to measure distortion.

In mathematical terms:

The Mean Squared Error (MSE) is calculated with the following equation: [8]

$$MSE = \frac{1}{mn} \sum_{i=0}^{m-1} \sum_{j=0}^{n-1} [I(i,j) - K(i,j)]^2 \quad Equation\ 1$$

Where:

    m and n are the image dimensions

    I(i,j) is the pixel value in the original image

    K(i,j) is the pixel value in the steganographic image

---

[8] https://en.wikipedia.org/wiki/Peak_signal-to-noise_ratio

PSNR = $10 * \log_{10} (MAX_I^2 / \text{MSE})$ [See Footnote 9]

$= 20 \log_{10} (MAX_I / \sqrt{\text{MSE}}$

$= 20 \log_{10} MAX_I - 10 \log_{10} \text{MSE}$

Where MAX_I is the maximum possible pixel value (255). (MAX_I is squared to be standard with the MSE which is squared (to remove the negative value.)

Decibels are used for measuring PSNR. The human perception of visual differences is measurable in decibels. A decibel difference of 0.5 to 1 is humanly perceptible. Also, the range of differences can be quite large. Using a logarithmic scale allows for a more compact representation.

A higher PSNR value indicates the steganographic image is more similar to the original cover image. A decibel range of 50-55 dB is considered good.

As the MSE decreases (indicating less difference between the 2 images), the PSNR increases.

## Structural Similarity Index Measure (SSIM) is another measure of distortion or

alteration of an image with a steganographic message or watermarked. SSIM is now considered a more superior measure than PSNR although PSNR is easier to computer. Basically, SSIM is a ratio of the differences of the average (mean, variance (standard deviation) and covariance) pixel values of the 2 images. (The original image and the stego or watermarked image.) It is calculated with the following equation:[10]

$$SSIM(x, y) = \frac{(2\mu_x\mu_y + c1)(2\sigma xy + c2)}{(\mu_x^2 + \mu_y^2 + c1)(\sigma_x^2 + \sigma_y^2 + c2)} \quad Equation\ 2$$

---

[9] https://en.wikipedia.org/wiki/Peak_signal-to-noise_ratio
[10] https://www.geeksforgeeks.org/performance-metrics-for-image-steganography/

Where:

μx and μy are the average pixel values of x and y

σx and σy are the variances of x and y

σxy is the covariance of x and y

c1 and c2 are constants to stabilize the division with weak denominators.

c1 = 0.02 and c2 = 0.03

Values close to 1 indicate similarity.

## 1.2. Steganographic Methods for Digital Images

Lempel Ziv Welch (LZW) is a compressed technique that is used in conjunction with other techniques to securely transmit medical images. (Magdy, et. al., 2022) This technique uses LZW compression, followed by encryption with AES, after which CEDE (Compressed Encrypted Data Embedding) is employed. LZW is a lossless compression technique that builds a dictionary of recurring data string sequences and replaces them with shorter codes. (Hamza, et. al. 2021)

The dictionary starts with an entry for all possible single characters (256 ASCII characters). The data is read character by character. Sequences of characters are derived. If a sequence is not in the dictionary, it is added and assigned a new code. The process continues, building longer and longer sequences

The dictionary is dynamic. As the compression process continues, the dictionary gets bigger, while allowing for more efficient compression of repetitive patterns

The lossless feature of LZW makes it attractive for use in medical images that require lossless processing for integrity. The compressed data can then be used in steganography as a payload to compare with the image for establishing originality.

This technique allows for the embedding of a compressed, encrypted copy of the image – a hash – within another image (carrier image). Thus, integrity and confidentiality can be maintained. Integrity is maintained as the compression is lossless. Confidentiality is maintained as the medical image is not sent in the clear. Only the carrier image is sent in the clear.

Histogram Reversible Data Hiding (HS-RDH) is another kind of steganography that does not change the values of the pixels directly. In terms of information hidden in medical images, one must also consider the medical knowledge itself. **Contrast Enhancement** can help better visualize the internal structure of organs so that ROIs, such as tumors can be more prominent. This provides more detailed analysis, ensures privacy, and better diagnosis. At the same time, this technique can embed information into a medical image, without distortion of the medical diagnosis. This is referred to as **Reversible Data Hiding with Contrast Enhancement (RDHCE).**

The process starts with making a histogram (bar chart) of the intensities of the pixels. This is followed by manipulating the bars to make or identify bars where information can be hidden. Equalizing the bars will improve contrast and adjust brightness. In a sense, this will bring out "hidden information" as the contrast will bring artifacts (e.g. tumors) to light (literally). Concurrently, the equalizing process makes bars in which information can be hidden without distorting the medical image.

One method to do this is to convert the color components from the Red Green Blue (RGB) color model to the Hue Saturation Value (HSV) color model (value being intensity/brightness). The human eye is least sensitive to red and most sensitive to blue. So, the RGB distribution is not uniform. HSV, however, is a color model that is closer to people's perception. The perceived color of an area can be defined by three components which are hue (H), saturation (S), and brightness or Value (V).  (Lee, et. al. 2023)

By converting the RGB image information into the model HSV then; manipulating the HSV model and hiding information in the V channel (intensity); followed by converting the HSV model back to RGB, secret information can be passed.

This is done by making a histogram of the RGB model and manipulating the range of intensity. This way hue and hue saturation are preserved obviating image distortion. The V channel is manipulated thus:

'H($pi$) represents the pixel value $pi$ in a gray scale image where $pi$ ∈ {0, 1, ..., 254, 255}. In the histogram of *pi*, the pixel values between the peaks $pp1$ and $pp2$ unchanged, and the outer pixels values are shifted so that each of the peaks can be split into two adjacent bins. This makes new bins suitable for data hiding.' (Lee, et. al. 2023)

Then, the HSV model is converted back to RGB with the hidden message now embedded.

With medical images, most of the background is black. Thus, the pixel distribution of the histogram tends to be concentrated at the peaks. A higher number of peaks can provide additional space to conceal confidential information, resulting in a larger embedding capacity. So, this technique is well suited to medical images. The image contrast enhancement not only improves the quality of the stego-image but also offers a significant hiding space.

It should be noted that this method is also well suited to colored medical images, not just black and white images.

Pixel Value Differencing (PVD) is a steganographic method developed by Wu, et. al. 2003.

The basic concept is to embed variable numbers of bits in different pixels because changes in pixels in smooth areas (areas without much color differences between adjacent pixels) are easily seen by the human eye than changes in rough areas (areas with large color differences between adjacent pixels). The cover image is divided into non-overlapping 2 pixel blocks. The blocks are categorized by severity of difference between. A small difference is a smooth block. A large difference is an edge block. (An edge being either a color or luminance boundary, where the value one side is high and the value on the other side is low.) More data is embedded in rough areas than smooth areas. Thereby, keeping the perceptibility of the embedded data to a minimum and imperceivable.

Pixel values range from 1-255 (0-254). The differences range from 0-254 and can be negative. The absolute differences are used. Categories are defined by powers of 2. Replacement bits are such that the difference between the pixels remains as close as possible.

To calculate the number of pixels that can be embedded and the difference between the pixel pairs, the following calculations are made:

Where *n* – the number of pixels that can be embedded;

    Ri – range of pixels differences defined as a number {1, 2, … 255}

    *d* – the difference between pixels

    *u* – upper limit

    *l* – lower limit

    *k* – index of range (of differences)

Given a 2 pixel block: $n = \log_2 (u_k - l_k + 1)$ The power, base 2, that can raised to, to equal the number of differences between the pixels, plus 1. Thus, *n* can range from 0-255.

Then, a substream *S*, of the data to be embedded is selected. It also has a length of *n* bits. Then, d' is calculated, for the substream *S*. Where $d' = l_k + b$, where $0 <= b <= u_k - l_k$ Since d and d' are in the same range of $u_k$ to $l_k$, then d can be replaced with d' with imperceptibility.

Pixel replacement is done with a function:

$f((g_i, g_{i+1}), m) = (g'_i, g'_{i+1}) = (g_i - ceiling_m, g_{i+1} + floor_m, if\ d\ is\ odd$

$\qquad\qquad\qquad = (g_i - floor_m, g_{i+1} + ceiling_m, if\ d\ is\ even$

Where *g* – grayscale value for a pixel; m = d' – d; $floor_m = \lfloor \frac{m}{2} \rfloor,\ ceiling_m = \lceil m/2 \rceil$

Using the ceiling and floor of the pixel difference values is a distortion reduction policy that equally distributes the distortion caused by the grayscale replacements. This makes the grayscale change within the block less perceptible.

This inverse process may cause d' to fall outside real boundaries. When this happens, the pixel block is ignored. This can be checked when extracting data and such blocks can also be ignored when extracting data. In practicality, this is not a common occurrence.

For example, in figure 3, the original grayscale values of 50 and 65, have a difference range of 15. The embedding categories k, step by power 2. This puts 15 in the range of 2^3 – 2^4 = 8 – 23. The 4 bits that can be embedded are a decimal 10. The lower boundary plus the embedded value is 18. Thus *m* = 3 (18-15, the difference between the original pixels and the new pixels with the embedded data). Floor(m/2) = 1. Ceil(m/2)=2. 50 – 2 = 48. 65 + 1 = 66. Which yields a new pixel pair of 48 and 66, with a difference of 18.

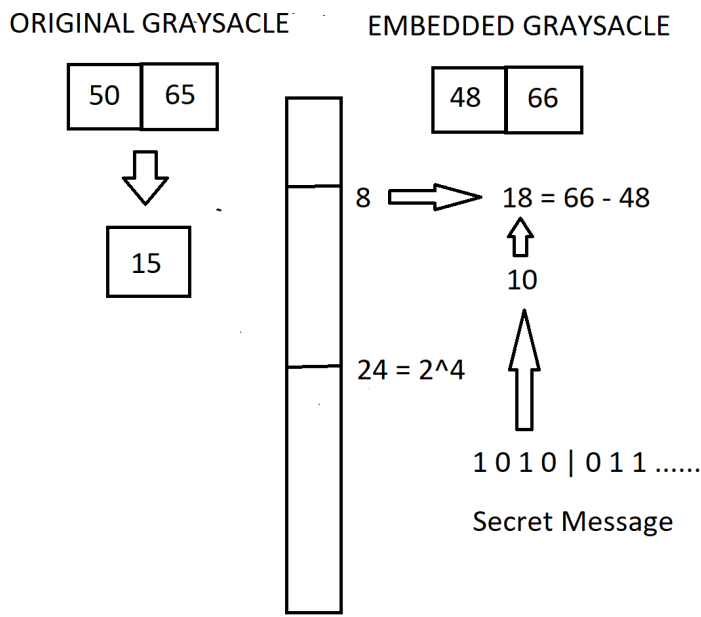

*Figure 3* Illustration of PVD embedding process. Courtesy Wu, et. al. (2003)

For extraction, the seed is reapplied for the same pseudorandom number for traversing the image. Blocks to be skipped can be calculated as above. Then, recalculated with the same equation only using the new differences. The proof is:

$$(g'_i, g'_{i+1}) = f((g_i + g_{i+1}), u_k - d)$$
$$= f((g_i + g_{i+1}), d^* + u_k - d - d^*)$$
$$= ff((g_i + g_{i+1}), d^* - d), u_k - d^*)$$
$$= f((g'_i + g'_{i+1}), u_k - d^*)$$

It is important to note that not only is the secret message retrieved but, the original image is restored by the extraction process. Such a process allows for embedding a watermark and its retrieval, while maintaining the integrity of the image. For medical imaging, this is an integral function. So, PVD can be employed with other techniques as outlined in this paper to mark and evaluate the transmission of medical images.

The author notes that the seed and random generator are shared secrets. In keeping with Kerckhoff's cryptographic principles (Auguste 1883), we can assume that everyone knows the random generator's algorithm. The seed is a symmetric key, which must be guarded and has all the problems of distribution that a symmetric key has. It should not be transmitted together with the stego image.

## 2. Watermarking

A watermark is a picture embedded in the host media so that it can't be separated from the data. Thus, by using watermarking, the work is still available but watermarked. Watermarks can't be separated from the image but do not detract / degrade from the image. (Magdy, et. al. 2022) Watermarking is distinct from steganography in that a watermark is meant to be seen or known— it is not hidden and is expected to be found. There is no hidden information with a watermark like there is in steganography and imperceptibility is not a goal with watermarking.

Akin to steganography, watermarking can be either in the spatial or frequency domain. Spatial watermarking has better capacity but is much less resistant to attacks. (Badshah, et. al. 2016) Which, can make the watermark useless.

Several watermarking techniques have been applied to medical images: reversible, irreversible, tamper detection and recovery; tamper localization and lossless recovery.

Heavy payloads (of a watermark) can cause image degradation. To counter this problem, compression (of the watermark) can be used.

Fragile Watermarking can be used to authenticate and prevent image degradation. This method watermarks blocks of data individually instead of the entire image at once. If the watermark in a block fails, the location of the failure is known. These watermarks are not intended to be "strong" and resist attacks (distortion). One of the proposed methods to do so is Singular Value Decomposition (SVD), which divides the image into blocks (8 x 8); then embedding authentication data – a key - in the LSB 'plane' (with the LSB technique) and then, calculating a modular value for the block. A second key is used to randomize the blocks. Both keys are needed to reverse the process. In reversing should the first key not match, the location of degradation is known. (Oktavia, et. al. 2004)

Watermarking with K-Means[11] is a method suggested by Kahla, et. al. 2023. This method has 4 steps:

1. Generate a binary hash from a watermark
2. Embed the watermark
3. Extract the watermark

---

[11] K-means clustering is an AI method that chooses points by using a centroid, akin to calculating the center of gravity. The calculations can be readjusted depending upon the grouping, number of members, balance, etc. These groupings give a balanced separation of entities into groups.
https://en.wikipedia.org/wiki/K-means_clustering

4. Evaluate the watermark for changes

In the first step, the watermark is converted into a binary 2D array. Using the K-Means method, group similar colors into clusters of the watermark into 8 clusters. Replace the clusters with their centrix. Convert the centrix to binary. Concatenate the clusters into a binary string. Generate a binary hash by vertically and horizontally replicating the concatenated binary string. Replace the LSB of the green channel with the watermark.

To extract, recreate the watermark from the LSB of the green channel.

To compare, XOR the recreated watermark with the original watermark. If altered, the results is 1. If unaltered, the result is 0.

Khala, et. al., 2023 describe test results of a PSNR 50 – 55 db and correctly locating "attacks", embedded distortion in the original image, such as text. What the author (this author) is unclear of, is although the watermark can clearly identify the location and severity of distortion in the original image, how is the integrity of the original image not disturbed by the watermarking process.

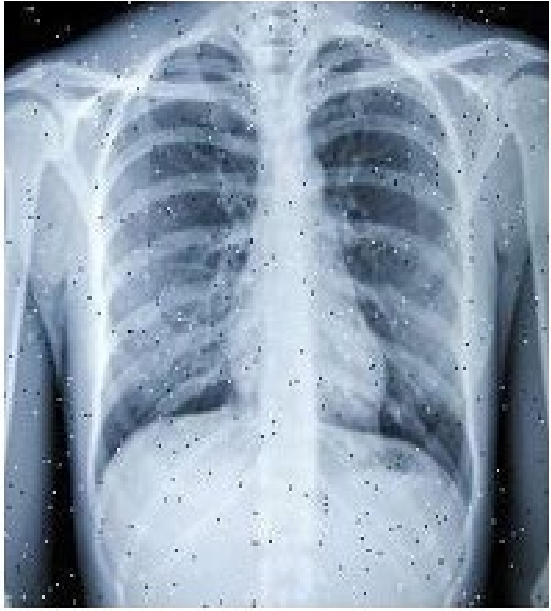
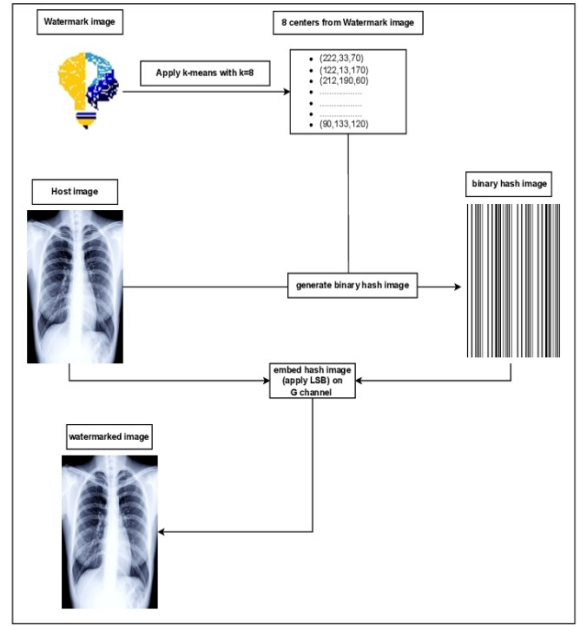
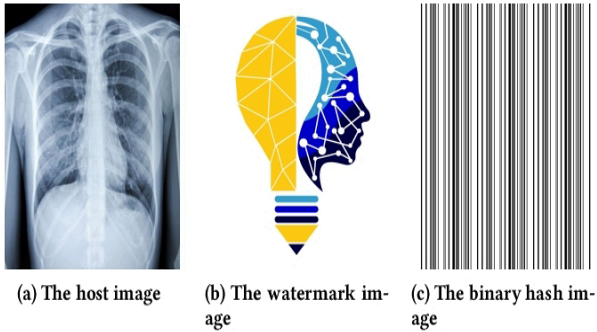
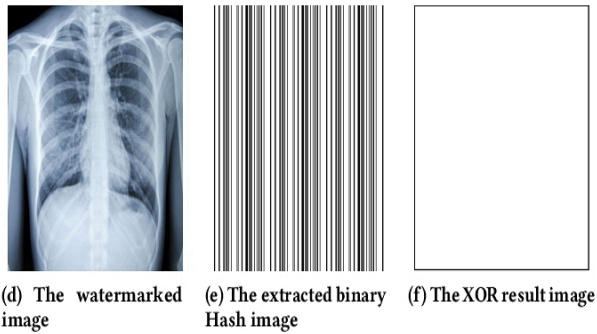
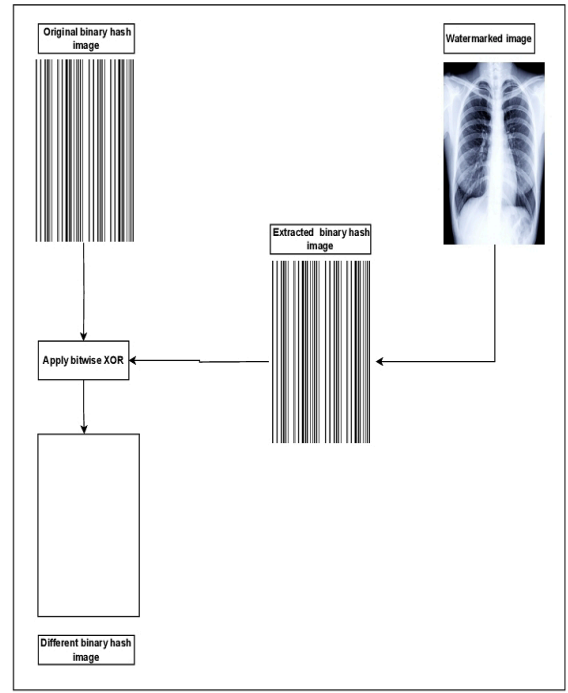

*Figure 4.* Top Left: Chest X-Ray. Top Right: Processing the Watermark and Producing a hash. Bottom Left: Recreating the watermark hash and extracting the watermark hash. Then, comparing the two. Bottom Right: Comparing the original hash to the extracted hash with an XOR. Pictures courtesy Kahla, et. al. (2023)

### Lossless Compression Watermarking with LZW.

Badshah, et. al. 2016 proposed a lossless watermarking technique using LZW compression. Badshah, et. al. noted that while there are many online security techniques such as MAC, digital signatures, etc., they are not suited for the sensitive data security that digital images need. While these methods can detect alterations, they cannot detect where in the data the alterations occurred. This bit-to-bit originality is necessary for medical images. Watermarking can provide recovery as well as authentication and integrity checking.

Badshah, et. al. 2016 worked with sonograms. Sonograms have clear ROIs. The rich picture is in the center of the image, while the outer regions, especially the top and sides are not of interest, RONIs. This makes embedding a watermark in a RONI, unlikely to affect image degradation. The watermark is an ROI with a key applied. Upon extraction, the key is applied, and the ROI extracted. The compression allows for the watermarked ROI to be compressed, so it can fit into the RONI without disturbing the ROIs.

The author notes, that while this is a very useful process, it is limited to this kind of data – sonograms or; similar kinds of data that have significant RONIs to support successfully applying the method.

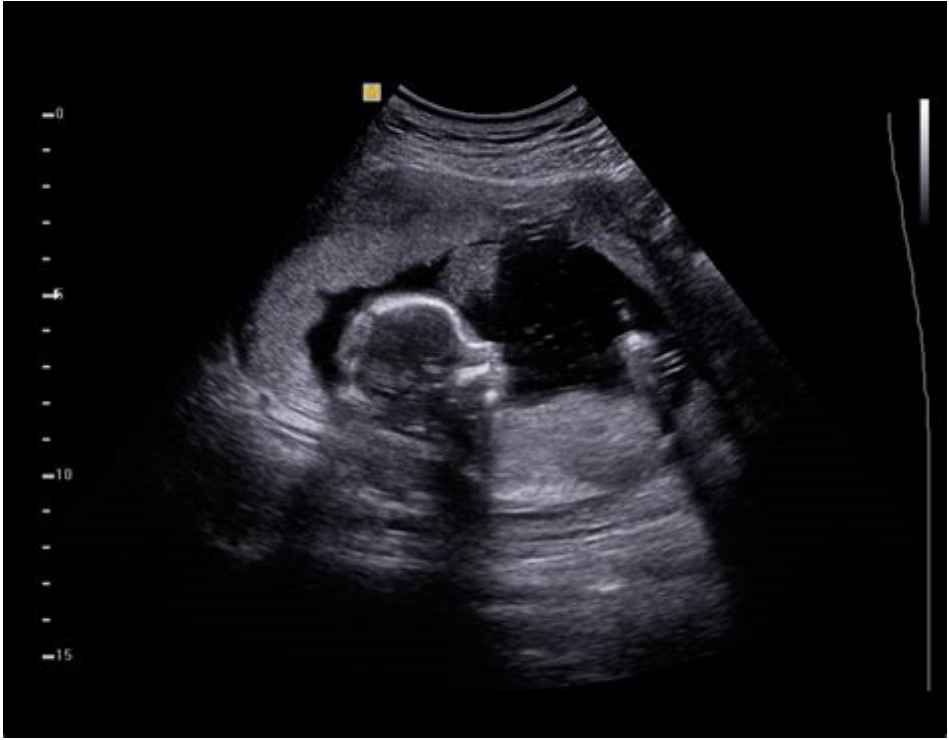

*Figure 4* Notice large black areas. These are RONIs – Regions of Non-Interest. Courtesy Pinterest Publicly available piicture of sonogram.

Singular Value Decompostion (SVD) – is another of watermarking first presented by

Oktavia et. al. 2004. An object in coordinate geometry can be represented by coordinates in a matrix. The object can be translated and transformed with matrix arithmetic. The process is a rotation, followed by a translation (shift), which is followed by a final rotation. One such transformation can be a linear transformation that will turn the object into a line. We learn from this that any object can be expressed by 3 matrixes operated on in the order of multiplication, addition and a final multiplication. This can be applied to image compression because the number of values needed for these calculations can be much less, as the size of the matrix of the original image grows.

The multiplication can be reduced to one column-row times one value from a diagonal of a matrix; added to one more column-row multiplication. When repeated several times, with successive column-row pairs, the addition of the resultant matrixes will produce the original image. The values in the matrixes being pixel values with a value from 0-255, representing one per pixel per byte or; a translation number.

The value to this approach is that fair fewer variables are needed to store the image data.

The SVD can be calculated for the image and embedded. When this is done, the original image can be reconstructed. The reconstructed image can be compared with the original image for verification and tamper detection.

Oktavia et. al. 2004 proposed to subdivide the image into 8 x 8 blocks, perform an SVD operation on each block and embed the SVD information as a watermark. This allows for the watermark to identify where the distortion occurred.

Any real matrix A = *m x n* can be expressed by 3 matrixes, A=USV$^T$. Matrixes U and V are ordinary matrixes. Matrix U is pseudo-orthogonal. The diagonal of matrix U has values and all the other elements are zero. The matrix is an identity matrix and these while referred to as singular values are also known as eigenvalues. U and V are not unique but the values of S'es diagonal, the eigenvalues are unique and determined by A. In effect, this is a linear transformation and another way of representing a matrix with a set of 3 matrixes. The utility of which, is the reduction in size of expressing the matrix A.[12] [13]

The savings of size is costs in computational time.

---

## 3. Conclusion

The author's previous research has already demonstrated that embedding an unknown, non-repeating, sequence (such as can be generated by use of the Fibonacci series) can uniquely, securely and indelibly watermark a digital image. Further research into if and how the use of LSB with a key produced by a random number generator would increase the difficulty of extracting the hidden image from the carrier could lead to a stronger method of steganography. This could be applied to any imaging requiring strong security—not just medical images. Integrity of the image can be maintained because the key information would simply be replaced with the original data, while the key being random, could not be just extracted by any attack. This would require, in addition to the key, a reversible process for embedding the key. Ex. XORing the significant bit. The probably of multiple attacks reversing the key image seems highly unlikely. While the author is unaware of such a method steganography, research might show that such a method is already in use.

Extending this concept to the integrity of medical images, a self-watermarking steganographic technique could create a pseudorandom number with a hash of the image. This would not provide confidentiality but, the addition of an injection vector, would. This hash could be transmitted with the image and then, upon receipt, extracted from the image, XORed with the image to reverse the image to its original state, then hash of the image could be recreated with the injection vector/key. The comparison of the 2 hashes would establish equivalency.

Another subject for further research is if using LSB in a chromatic steganographic encryption, the payload could be diffused between all 3 color channels without distorting the color of the image.

An encrypted or randomized SVD watermark seems the most appropriate measure as it distinguishes where exactly the loss of integrity was. If the distorted regions are RONIs, then the image can still be used and retransmission is not necessary.

## Future Research

The author is quite puzzled by these proposed, complicated steganographic schemes for the integrity of medical imaging. There are other methods of steganography that are well known that are easy to decipher, out-of-the-box, open source, in combination with the techniques described above—embedding a hash of the image (preferably an encrypted or randomized SVD); while not necessarily providing confidentiality, that could be used to ensure the integrity of an image and have no image degradation at all. If desired, a message authentication (MAC) could be employed. Then, standard encryption such as AES could be applied to provide confidentiality. The author finds the lack of such approaches puzzling.  Such a codec can be developed and tested for efficiency and efficacy. Research into which methods are currently being employed by medical imaging companies and why simpler methods of identifying medical images are not used should be studied.